\documentclass[amsmath,superscriptaddress,aps,prb,twocolumn]{revtex4-1}
\usepackage[colorlinks,linkcolor=blue,anchorcolor=blue,urlcolor=blue,citecolor=blue]{hyperref}
\usepackage{amsmath}
\usepackage{amssymb}
\usepackage{graphicx}
\usepackage{color}
\usepackage{bm}
\usepackage{float}

\begin{document}

\title{Anomalous non-Hermitian skin effect: the topological inequivalence of skin modes versus point gap}

\author{Gang-Feng Guo}
\affiliation{Lanzhou Center for Theoretical Physics, Key Laboratory of Theoretical Physics of Gansu Province, Lanzhou University, Lanzhou $730000$,
China}

\author{Xi-Xi Bao}
\affiliation{Lanzhou Center for Theoretical Physics, Key Laboratory of Theoretical Physics of Gansu Province, Lanzhou University, Lanzhou $730000$,
China}

\author{Han-Jie Zhu}
\affiliation{Beijing National Laboratory for Condensed Matter Physics, Institute of Physics, Chinese Academy of Sciences, Beijing $100190$, China}

\author{Xiao-Ming Zhao}
\affiliation{Department of Physics and Institute of Theoretical Physics, University of Science and Technology Beijing, Beijing 100083, China}

\author{Lin Zhuang}
\affiliation{State Key Laboratory of Optoelectronic Materials and Technologies, School of Physics, Sun Yat-Sen University, Guangzhou 510275, China}

\author{Lei Tan}
\email{tanlei@lzu.edu.cn}
\affiliation{Lanzhou Center for Theoretical Physics, Key Laboratory of Theoretical Physics of Gansu Province, Lanzhou University, Lanzhou $730000$,
China}
\affiliation{Key Laboratory for Magnetism and Magnetic Materials of the Ministry of Education, Lanzhou University, Lanzhou $730000$, China}

\author{Wu-Ming Liu}
\email{wliu@iphy.ac.cn}
\affiliation{Beijing National Laboratory for Condensed Matter Physics, Institute of Physics, Chinese Academy of Sciences, Beijing $100190$, China}

\begin{abstract}
\noindent Non-Hermitian skin effect, the localization of an extensive number of eigenstates at the ends of the system, has greatly expanded the frontier of physical laws. It has long been believed that the present of skin modes is equivalent to the topologically nontrivial point gap of complex eigenvalues under periodic boundary conditions, and vice versa. However, we find that this concomitance can be broken, i.e., the skin modes can be present or absent whereas the point gap is topologically trivial or nontrivial, respectively, named anomalous non-Hermitian skin effect. This anomalous phenomenon arises when the unidirectional hopping amplitudes leading to the decoupling-like behaviors among subsystems are emergence. The emergence of the anomalous non-Hermitian skin effect is accompanied by the mutations of the open boundary energy spectrum, whose structure exhibits the multifold exceptional point and can not be recovered by continuum bands. Moreover, an experimental setup using circuits is proposed to simulate this novel quantum effect. Our results reveal the topologically inequivalent between skin modes and point gap. This new effect not only can give a deeper understanding of non-Bloch theory and the critical phenomenon in non-Hermitian systems, but may also inspire new applications such as in the sensors field.
\end{abstract}

\maketitle

\begin{figure}
\includegraphics[width=8.5cm,height=6.3cm]{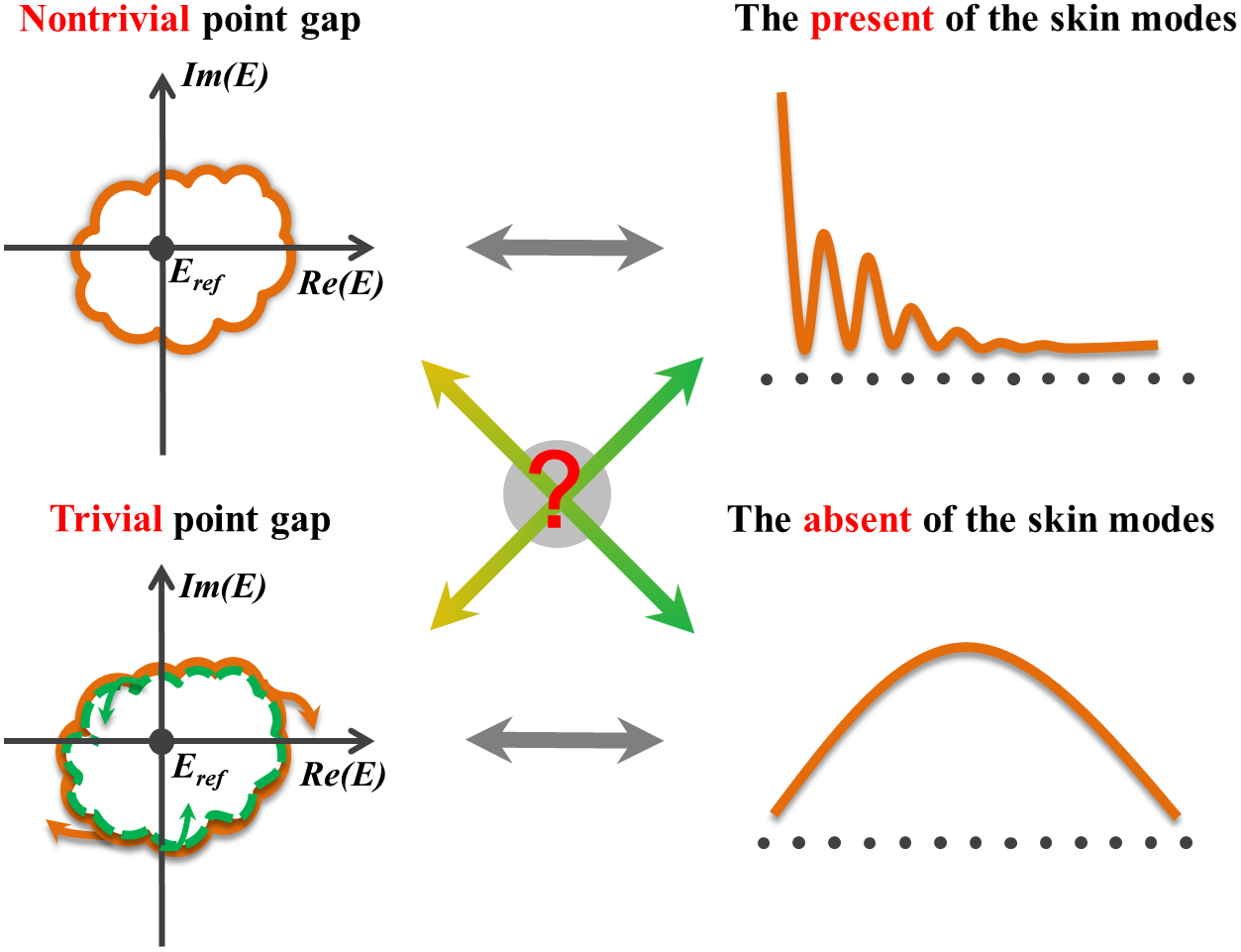}
\caption{\textbf{The relationship between the point gap and the non-Hermitian skin effect}. It takes for granted that the topologically nontrivial (trivial) point gap is equivalent to the present (absent) of skin modes. However, is this really a universal conclusion? In our work, we exhibit a new consequence of the anomalous non-Hermitian skin effect: the open boundary eigenstates being localized for the topologically trivial point gap while extended for the nontrivial point gap.}
\label{fig1}
\end{figure}

\noindent Recent theoretical and experimental studies on non-Hermitian systems have revealed many interesting phenomena without Hermitian counterparts \cite{Fran1, Xue2, Abba3, Li4, Wang5, Song6, Vecs7, Okug8, Yosh9, Guo10, Kawa11, Zhu12, Li13}. One of the quintessential features of non-Hermitian systems is the skin modes \cite{Yao14, Yok15, Zhang16, Bert17, Tran18, Lee19, Obse20, Yang21}, namely, the accumulation of the majority of eigenstates at the boundaries of the system. The non-Hermitian skin effect will cause the nullities of the conventional bulk boundary correspondence \cite{Yao14}. Specifically, the eigenvalues under different boundary conditions have a crucial distinction. The non-Bloch band theory is then proposed, from which the generalized Brillouin zone and the continuum bands can be obtained to redefine the topological invariant and reproduce the open boundary band structures \cite{Yao14}$^{,}$ \cite{Yok15}. The exotic accumulation phenomenon of the eigenstates has important applications in topological sensors \cite{Budi22}$^{,}$ \cite{McDo23}, the integrated optical chip \cite{Seba24}, and topological lasing \cite{Long25}$^{,}$ \cite{Zhu26}. So far, it is widely believed that the presence of the non-Hermitian skin effect is tantamount to the topologically nontrivial point gap \cite{Zhang16, Kawa27, Okum28, Gon29, Denn30, Yi31}and vice versa. which can be explained as follows. The Hamiltonian under open boundary conditions is always topologically trivial for the point gap. Thus, if the Hamiltonian under periodic boundary conditions is nontrivial, the non-Hermitian skin effect is present inevitably [Fig. \ref{fig1}].

\begin{figure*}
\includegraphics[width=18cm,height=6cm]{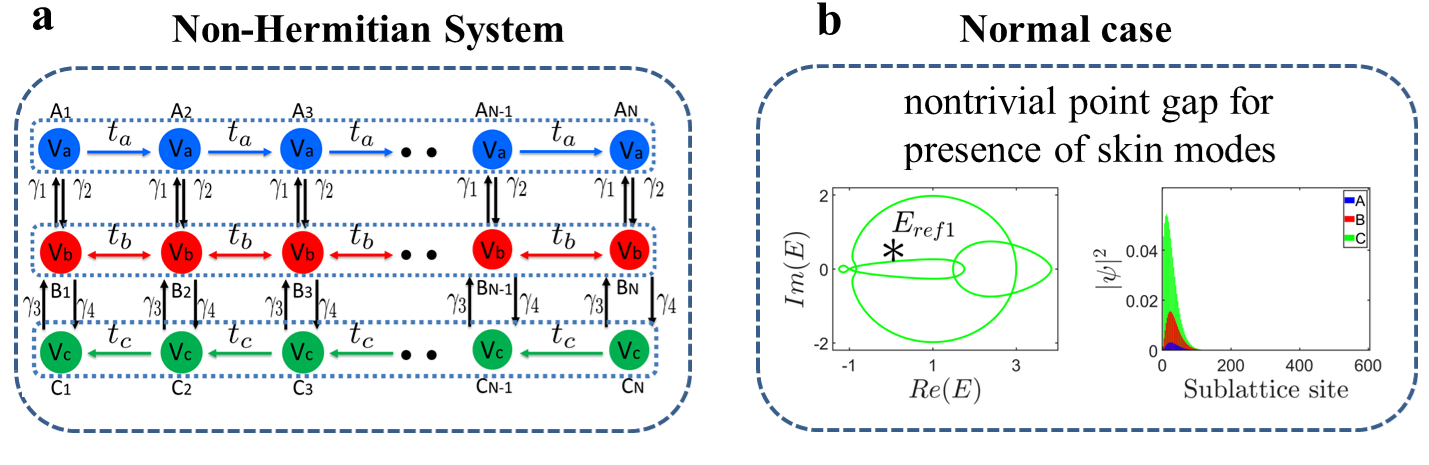}
\includegraphics[width=17.5cm,height=11cm]{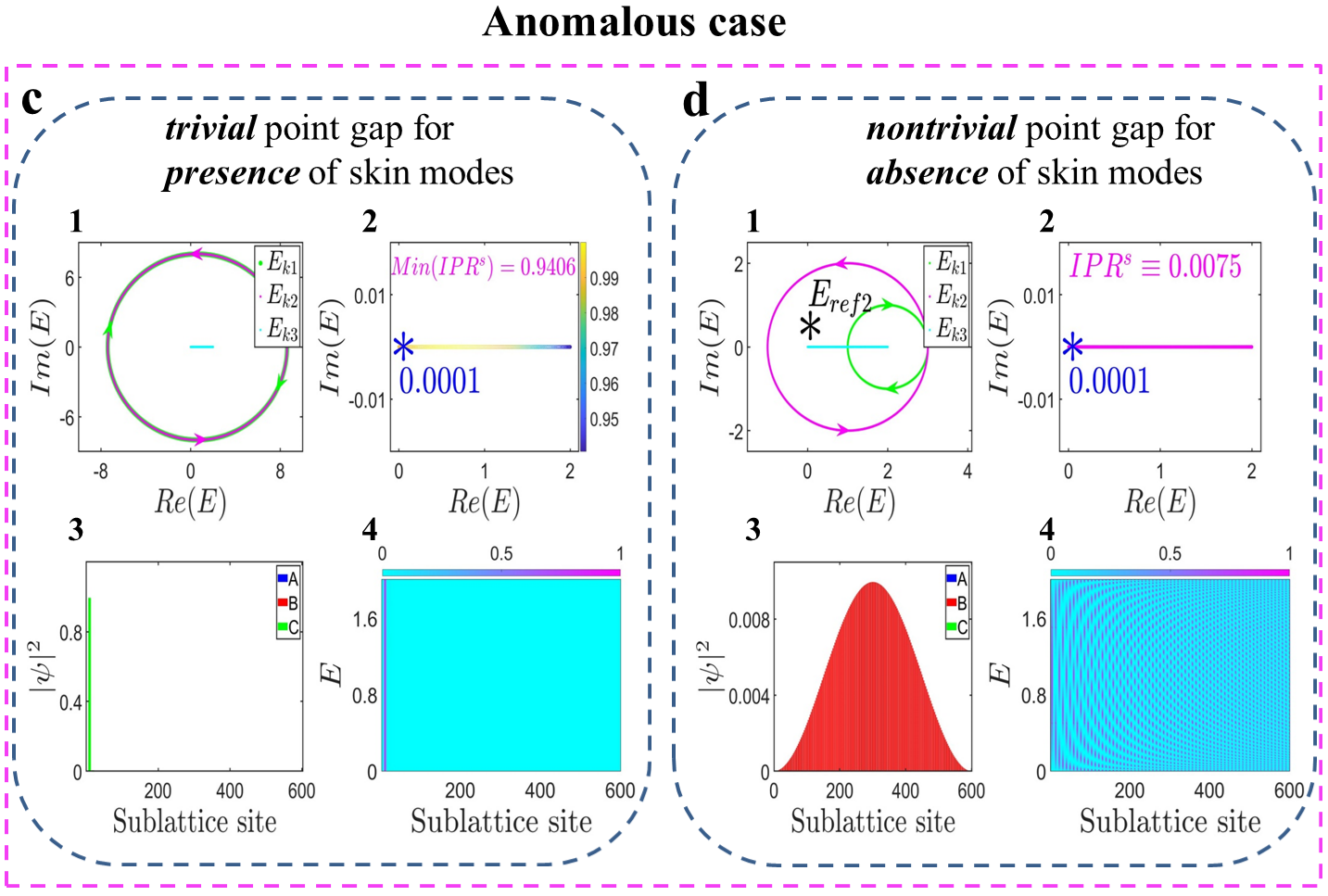}
\caption{\textbf{The anomalous non-Hermitian skin effect}. \textbf{a} Schematic representation of our system. \textbf{b} The normal case where the topologically nontrivial point gap stands for the occurrence of the non-Hermitian skin modes.  The reference point is $E_{ref1}=\frac{1}{2}i$. The parameters are $N=200$, $t_{a}=1$, $t_{b}=\frac{4}{5}$, $t_{c}=2$, $\gamma_{1}=\gamma_{2}=1$, $\gamma_{3}=\frac{1}{10}$, $\gamma_{4}=\frac{1}{2}$, $V_{a}=2$ and $V_{b}=V_{c}=1$. \textbf{c1} The momentum space spectra, in which $E_{k1}$ and $E_{k2}$ will display two circles with the same radius but the opposite rotation direction when $k$ changes from $0$ to $2\pi$. Thus, the point gap is topologically trivial for any reference point. \textbf{c2} The open boundary eigenvalues. Different colors are different values of $IPR^{s}$, which is finite for the corresponding eigenstate. \textbf{c3} The distribution of the open boundary eigenstate, which is localized, rather than extended. \textbf{c4} The normalized eigenstates as a function of eigenvalues. All eigenstates are localized at the boundaries of the system. \textbf{d1} The nontrivial point gap obtained from $(Re[det[H(k)]], Im[det[H(k)]])$ as $k$ varies from $0$ to $2\pi$ for the reference point $E_{ref2}=\frac{1}{2}i$. \textbf{d2} The open boundary eigenvalues colored by their respective $IPR^{s}$, being a constant approaching zero. \textbf{d3} The distribution of the open boundary eigenstate, in which only the occupation on the $B$ sub-chain is non-zero, while both the $A$ sub-chain and the $C$ sub-chain have no occupation. \textbf{d4} The normalized eigenstates with the change of eigenvalues. All eigenstates are extended, rather than localized, on the open chain. The parameters for \textbf{c} and \textbf{d} are the same as Fig. \ref{fig55}a and Fig. \ref{fig55}b in \textbf{Methods}, respectively.}
\label{fig2}
\end{figure*}

However, we uncover that the open boundary eigenstates can exhibit the anomalous non-Hermitian skin effect [Fig. \ref{fig2}], where skin modes can be ever present even though the point gap is trivial, while skin modes are absent for the nontrivial point gap. This phenomenon occurs as long as the unidirectional hopping amplitudes leading to decoupling-like behaviors of the system are considered. Following the appearance of the inequivalence between skin modes and point gap, the open boundary eigenvalues will produce a sudden change. Specifically, the multifold exceptional points, whose degree of degeneracy is proportional to the system size, will be displayed in the open boundary eigenvalues which have significant differences compared with the continuum bands. The physical properties of the eigenvalues may have applications in the sensors field. Finally, an experimental setup with electric circuits is implemented to realize our system.

\noindent $\textbf{Results}$

\noindent We consider a non-Hermitian three-band system (Fig. \ref{fig2}a), whose Hamiltonian reads as
\begin{flalign}
H=&\sum\limits_{n=1}^{N}\Big[t_{a}{C^\dag}_{A,n+1}{C}_{A,n}+t_{c}{C^\dag}_{C,n}{C}_{C,n+1}+\notag\\
&t_{b}({C^\dag}_{B,n}{C}_{B,n+1}+{C^\dag}_{B,n+1}{C}_{B,n})+\gamma_{1}{C^\dag}_{A,n}{C}_{B,n}\notag\\
&+\gamma_{2}{C^\dag}_{B,n}{C}_{A,n}+\gamma_{3}{C^\dag}_{B,n}{C}_{C,n}+\gamma_{4}{C^\dag}_{C,n}{C}_{B,n}+\notag\\
&V_{a}{C^\dag}_{A,n}{C}_{A,n}+V_{b}{C^\dag}_{B,n}{C}_{B,n}+V_{c}{C^\dag}_{C,n}{C}_{C,n}\Big],
\label{1}
\end{flalign}
where ${C^\dag}_{A,n}$ (${C}_{A,n}$), ${C^\dag}_{B,n}$ (${C}_{B,n}$) and ${C^\dag}_{C,n}$ (${C}_{C,n}$) represent the creation (annihilation) operators for sublattice $A$, $B$ and $C$ in $n$-th unit cells. For $A$ sub-chain, only right hopping $t_{a}$ is considered. Conversely, only left tunneling $t_{c}$ is considered for the $C$ sub-chain. Meanwhile, the $B$ sub-chain is Hermitian with the hopping amplitude being $t_{b}$. $\gamma_{1}$ and $\gamma_{2}$ stand for the non-Hermitian hopping between $A$ and $B$ sub-chain, while $\gamma_{3}$ and $\gamma_{4}$ are the non-Hermitian hopping between $B$ and $C$ sub-chain. $V_{a}$, $V_{b}$ and $V_{c}$ stand for the on-site potential for sub-chain $A$, $B$ and $C$, respectively. Without loss of generality, we take all parameters as real numbers.

To clarify our results unequivocally, we start with the normal case for comparison. The non-Hermitian skin effect can be generally confirmed by the nontrivial point gap, or equivalently, by non-zero spectral winding $W=\frac{1}{2\pi i} \int_{0}^{2\pi}dk\partial_{k}\ln\det[H(k)-E_{ref}]$ \cite{Zhang16} $^{,}$ \cite{Okum28, Gon29, Denn30, Yi31}, in which $E_{ref}$ is an arbitrary reference point. Such as, we can choose $E_{ref}=\frac{1}{2}i$ and $W=1$ is received (see $\textbf{Methods}$ and Supplementary Information\cite{ennd32} for more details). Thus, the system holds the non-Hermtian skin modes, as shown in Fig. \ref{fig2}b.

\noindent $\textbf{The anomalous presence of skin mode for}$
$\textbf{the trivial point gap}$

\noindent  We now explore the localization properties of the open boundary eigenstates under the unidirectional hopping amplitudes ($\gamma_{1}=\gamma_{3}=0$). To confirm that the spectral winding number is zero, corresponding to the trivial point gap, the conditions of Hamiltonian \eqref{1} will be further restricted as $V_{a}=V_{c}$, $t_{a}=t_{c}$, i.e., the generalized Brillouin zone is a unit circle \cite{ennd32}. In addition to the quantitative calculations of spectral windings, the topological properties of the point gap also can be gained intuitively. Namely, if the area of the curve surrounded by ($Re[det[H(k)]]$, $Im[det[H(k)]]$) is zero, the non-Hermitian skin modes will be vanishment and vice versa \cite{Zhang16}$^{,}$ \cite{Okum28, Gon29, Denn30, Yi31}. For our system, three eigenvalue branches in momentum space satisfy $|E_{k1}|=|V_{a}+t_{a}e^{-ik}|=|E_{k2}|=|V_{c}+t_{c}e^{ik}|$ and $E_{k3}=V_{b}+2t_{b}\cos(k)$, i.e., $E_{k1}$ and $E_{k2}$ will display two circles with the same radius but the opposite rotation direction when $k$ changes from $0$ to $2\pi$ [Fig. \ref{fig2}c1]. Therefore, for any reference point $E_{ref}$, the spectral winding is equal to zero exactly, and it seems that the non-Hermitian skin effect is not existence.

However, we find that this conclusion is not suitable for our non-Hermitian three-band system when the unidirectional hopping amplitudes are considered. To characterize the localization properties of all wave functions quantitatively, the inverse participation ratio ($IPR$) can be introduced as $IPR^{s}= \big(\sum\limits_{i=1} |\Psi_{i}^{s}|^{4} \big)/ \big(\sum\limits_{i=1} |\Psi_{i}^{s}|^{2}\big)^{2}$ for the $sth$ eigenstate $|\Psi^{s}\rangle$ \cite{Li33}$^{,}$ \cite{Li34}. In large system size, $IPR^{s}$ is finite for localized eigenstate, while it approaches zero for extended eigenstate. As shown in Fig. \ref{fig2}c2, $\min(IPR^{s})=0.9406$ as the energy value changes under numerical simulation, which means that all eigenstates are localized, rather than extended, even if the spectral winding number is constantly equal to zero. Here, the information about the exceptional points [discussed next] and the corresponding eigenstates have been excluded. We can arbitrarily present the distribution of an open boundary eigenstate in Fig. \ref{fig2}c3. It is obviously localized at the left boundaries of the system. Furthermore, Fig. \ref{fig2}c4 exhibits the distribution of all open boundary eigenstates associated with eigenvalues, where eigenstates have been normalized. Globally, all open boundary eigenstates are pinned at the left boundaries of the system, i.e., the non-Hermitian skin effect is present unexpectedly.

\noindent $\textbf{The anomalous absence of skin modes for}$
$\textbf{the nontrivial point gap}$

The non-Hermitian skin effect can be unexpectedly absent with the unidirectional hopping amplitudes of $\gamma_{1}=\gamma_{4}=0$, even though the point gap is topologically nontrivial. In this case, three eigenvalue branches in momentum space are $E_{k1}=V_{a}+t_{a}e^{-ik}$, $E_{k2}=V_{c}+t_{c}e^{ik}$ and $E_{k3}=V_{b}+2t_{b}\cos(k)$. The complex energy spectra $E_{k1}$ ($E_{k2}$) will rotate clockwise (counterclockwise) on the complex plane, and $E_{k3}$ just forms a line as $k$ varies from $0$ to $2\pi$. Therefore, as long as $V_{a}\neq V_{c}$ or $t_{a}\neq t_{c}$, there must exist a reference point $E_{ref}$ (Here we can choose $E_{ref2}=\frac{1}{2}i$) being surrounded only once and the area of the curve must be nonzero \cite{Yi31}, as shown in Fig. \ref{fig2}d1. Hence, it seems that our system holds the non-Hermitian skin effect.

However, we also can examine the localization properties of the open boundary eigenstates from the aspects of $IPR^{s}$. As shown in Fig. \ref{fig2}d2, $IPR^{s}\equiv0.0075$, which indicates that all eigenstates should be extended, not localized on the open chain. One eigenvalue in Fig. \ref{fig2}d2 can be selected to exhibit the distribution of the eigenstates. As shown in Figs. \ref{fig2}d3, the open boundary eigenstates are indeed extended only on the $B$ sub-chain, which conforms to the vanishment of the non-Hermitian skin effect numerically. Physically, $A$ ($C$) sub-chain has only unidirectional hopping to sub-chain $B$, but the reverse does not happen when $\gamma_{1}=0$ ($\gamma_{4}=0$), and $B$ sub-chain is Hermitian, which can be solved analytically \cite{ennd32}. Hence, there is no probability distribution on the sub-chain $A$ ($C$), and Bloch waves will be survived in the $B$ sub-chain. We further plot the distribution of all open boundary eigenstates corresponding to eigenvalues in Fig. \ref{fig2}d4. There is no large number of eigenstates localized at the boundary of the system, i.e., the non-Hermitian skin effect is absent whereas the point gap is nontrivial.

\begin{figure}[!htbp]
\includegraphics[width=4.27cm,height=3.24cm]{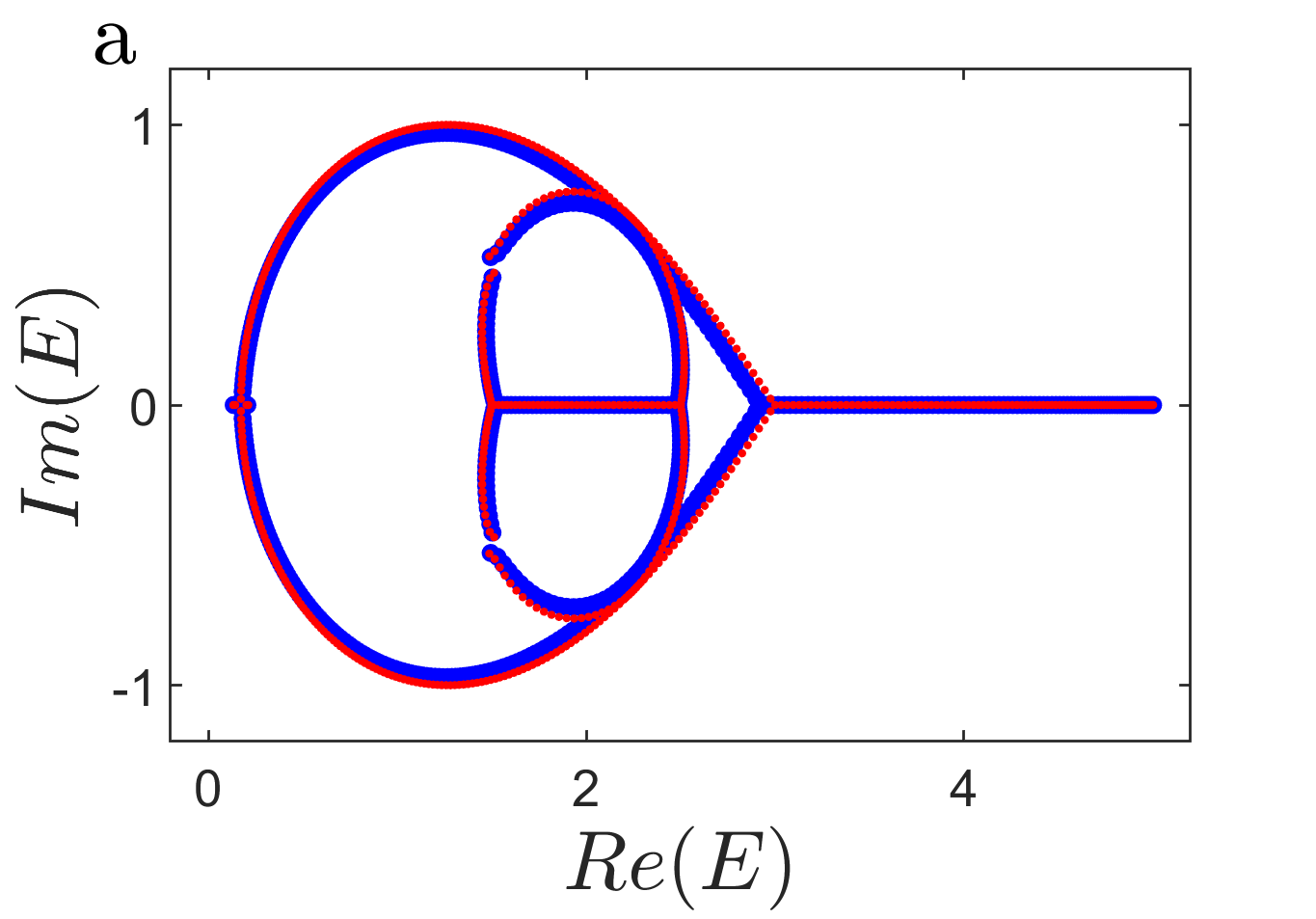}
\includegraphics[width=4.27cm,height=3.24cm]{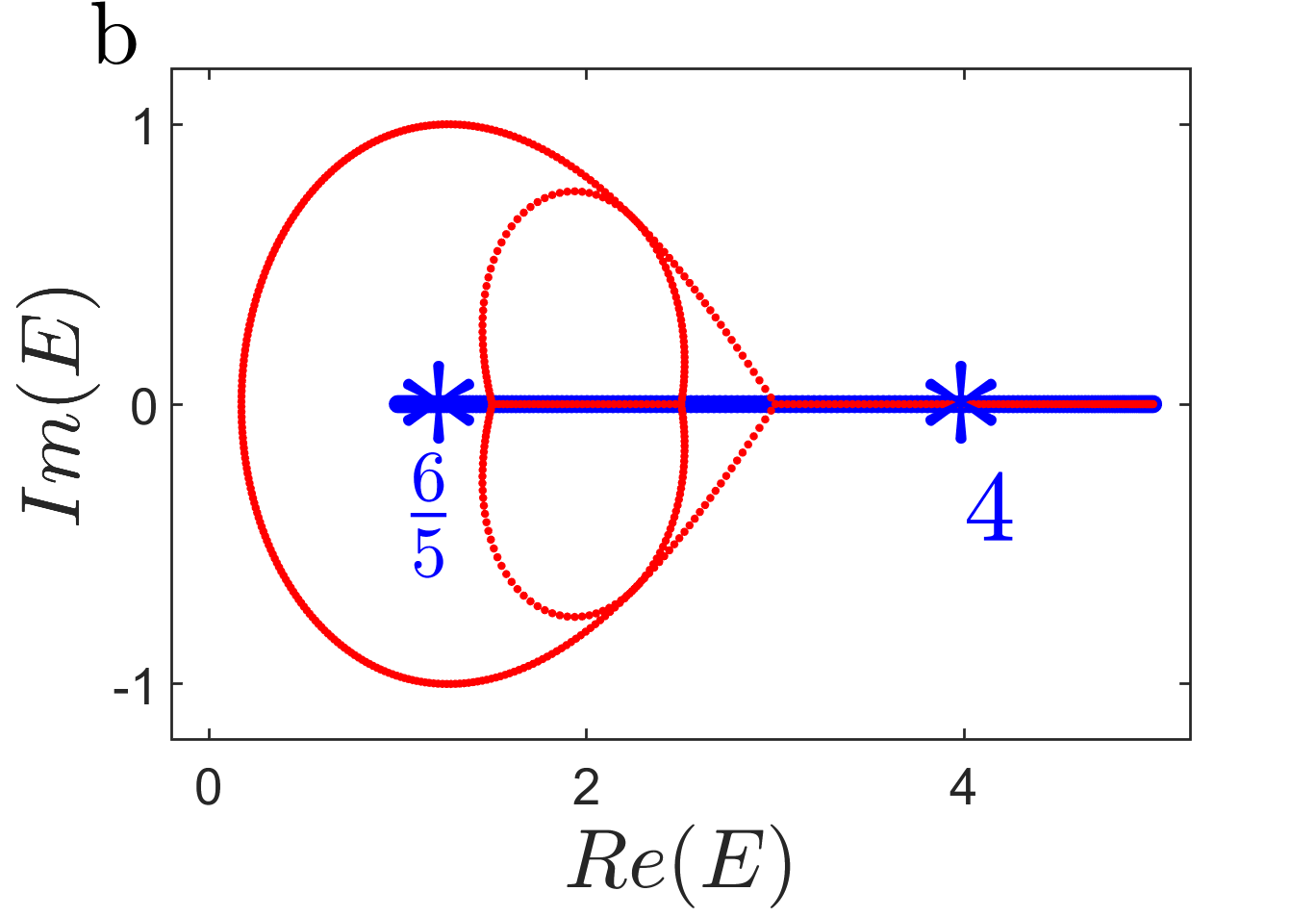}

\includegraphics[width=4.27cm,height=3.24cm]{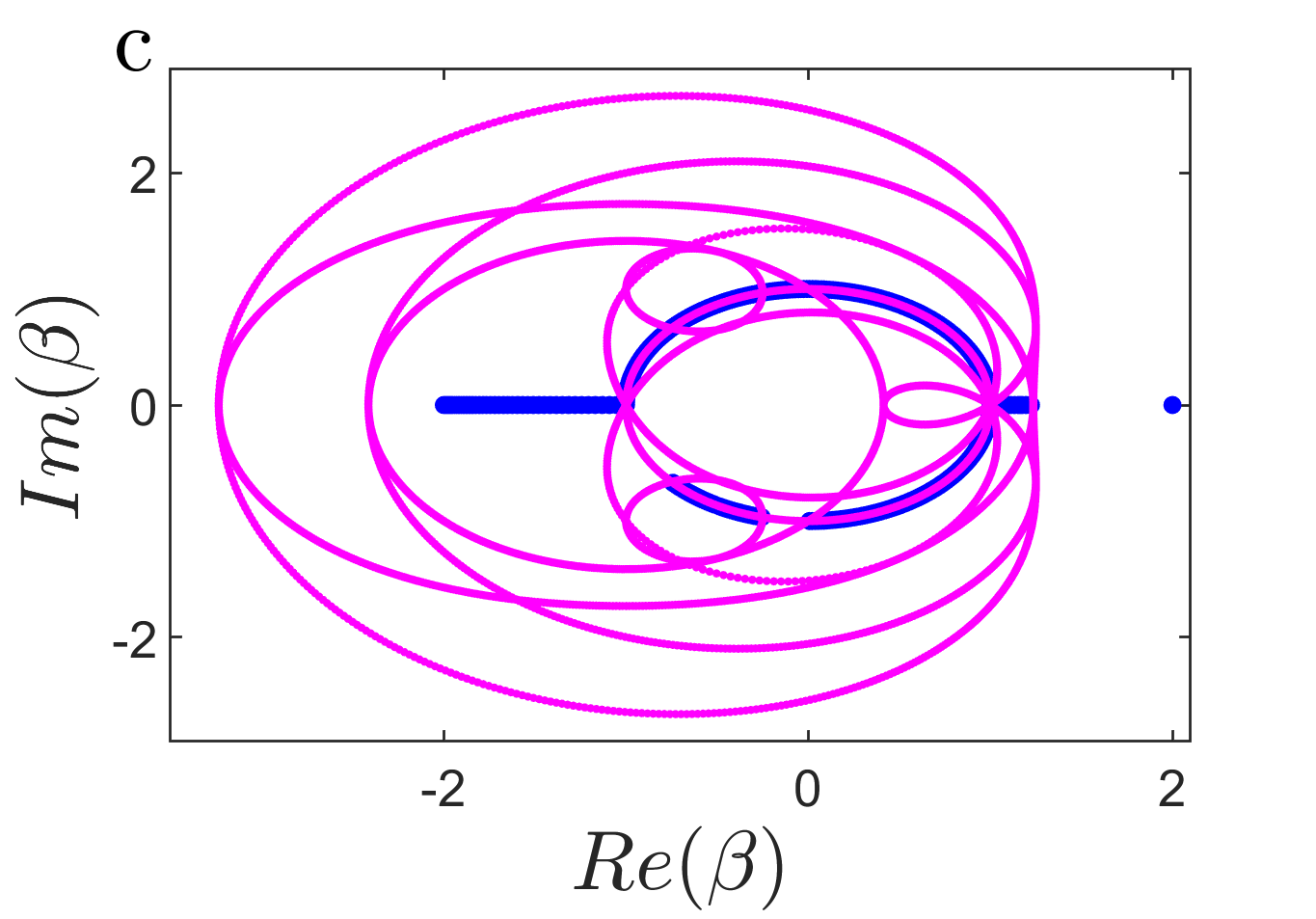}
\includegraphics[width=4.27cm,height=3.24cm]{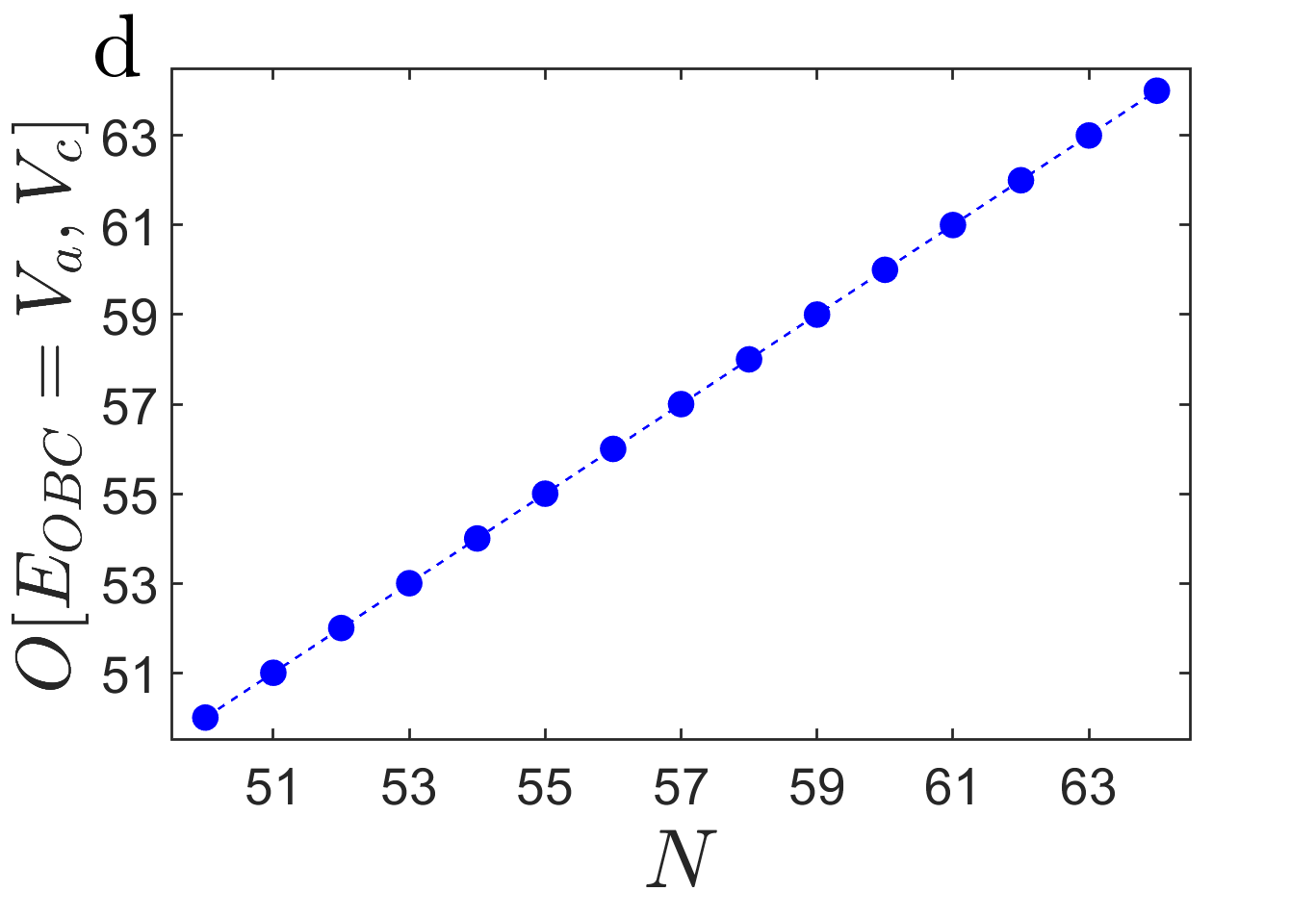}
\caption{$\textbf{The mutations of open boundary eigenvalues}$. \textbf{a} and \textbf{b} The energy spectra under open boundary conditions (the blue curve) and the continuum bands (the red curve). \textbf{b} After bringing two open boundary eigenvalues into $f(\beta, E)$, one energy value obtains $|\beta_{2}|=|\beta_{3}|$ ($E_{OBC}=4$) while the other corresponds to $|\beta_{1}|=|\beta_{2}|$ ($E_{OBC}=\frac{6}{5}$). \textbf{c} Auxiliary generalized Brillouin zone (the red curve) containing all information of $|\beta_{m}|=|\beta_{n}|$ ($\{m, n\}=\{1, 2, 3, 4\}$) with $f(\beta, E)=0$, and the generalized Brillouin zone obtained from the open boundary eigenvalues (the blue curve). \textbf{d} The number of certain open boundary eigenvalues versus the system size. \textbf{a} $\gamma_{2}=\frac{1}{500}$ and $\gamma_{3}=\frac{1}{400}$. \textbf{b}-\textbf{d} $\gamma_{2}=\gamma_{3}=0$. The rest parameters are $t_{a}=2$, $t_{b}=1$, $t_{c}=\frac{1}{2}$, $\gamma_{1}=1$, $\gamma_{4}=\frac{1}{10}$, $V_{a}=1$ $V_{b}=3$ and $V_{c}=2$.}
\label{fig3}
\end{figure}

\noindent $\textbf{The mutations of open boundary eigenvalues}$

\noindent Now, we analyze the properties of the energy spectrum where anomalous non-Hermitian skin effect occurs. According to the Hamiltonian \eqref{1}, the characteristic polynomial of our system is \cite{ennd32}

\begin{flalign}
f(\beta,E)=a_{2}\beta^{2}+a_{1}\beta^{1}+a_{0}\beta^{0}+a_{-1}\beta^{-1}+a_{-2}\beta^{-2}=0.
\label{212}
\end{flalign}

Obviously, the characteristic polynomial $f(\beta, E)$ is a quartic equation. The solutions can be numbered as
$\left|\beta_{1}\right|\leq\left|\beta_{2}\right|\leq\left|\beta_{3}\right|\leq\left|\beta_{4}\right|$ for the given E and $\left|\beta_{2}\right|=\left|\beta_{3}\right|$ is necessary to determine the generalized Brillouin zone and the derivative continuum bands. As shown in Fig. \ref{fig3}a, the open boundary energy spectrum can be well recovered by the continuum bands with $\gamma_{2}=\frac{1}{500}$ and $\gamma_{3}=\frac{1}{400}$.

However, the situation will have remarkable mutations when $\gamma_{2}=\gamma_{3}=0$, corresponding to the presence of the anomalous non-Hermitian skin effect. As shown in Fig. \ref{fig3}b, there is very little change for the continuum bands, while the open boundary eigenvalues instead will produce mutations and lay on the real axis. Hence, the open boundary energies and the continuum bands will not be compatible with each other. Further, it takes for granted that if every open boundary eigenvalue is brought into the characteristic polynomial $f(\beta, E)$, and the second and third largest $|\beta|$ are selected, the data should belong to the generalized Brillouin zone obtained from the non-Bloch band theory \cite{Yao14, Yok15, Zhang16, Yang21}$^{,}$ \cite{okom35}. However, unlike the established scenarios, this rule is nullified in the singular phenomenon. Detailedly, we can choose an open eigenvalue in Fig. \ref{fig3}b, such as $E_{OBC}=4$, and bring it into the characteristic polynomial $f(\beta, E)$ of our system. After calculation, one can receive $\big\{|\beta_{1}|, |\beta_{2}|, |\beta_{3}|, |\beta_{4}| \big \}_{E=4}=\big \{0.6667, 1.000, 1.000, 4.000 \big \}$, i.e., $|\beta_{2}|=|\beta_{3}|$. This equality is exactly the condition of obtaining the generalized Brillouin zone and the corresponding continuum bands. Therefore, $E_{OBC}=4$ can be reproduced by the continuum bands based on the non-Bloch band theory \cite{okom35}. Yet, another open boundary eigenvalue $E_{OBC}=\frac{6}{5}$ can be considered as well. Similarly, $\big\{|\beta_{1}|, |\beta_{2}|, |\beta_{3}|, |\beta_{4}| \big \}_{E=\frac{6}{5}}=\big \{1.000, 1.000, 1.600, 10.00 \big \},$ i.e., $|\beta_{2}| \neq |\beta_{3}|$ but $|\beta_{1}|=|\beta_{2}|$. Evidently, this equality is in contradiction to the condition of obtaining the generalized Brillouin zone, i.e., this open boundary eigenvalue must not belong to the continuum bands determined by $|\beta_{2}|= |\beta_{3}|$. Fig. \ref{fig3}b visually confirms those results.

\begin{figure*}
\includegraphics[width=15cm,height=6cm]{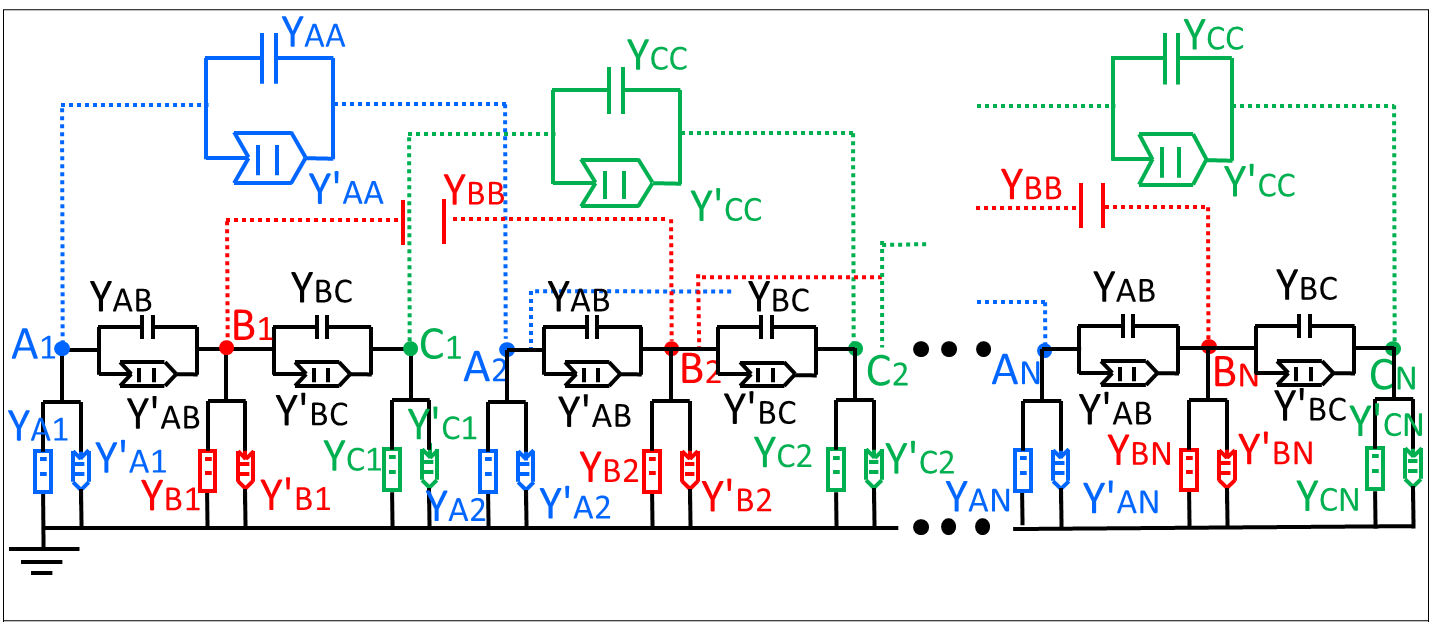}
\caption{\textbf{Schematics of the electric circuit configurations for realizing the Hamiltonian \eqref{1}}. The electric elements contain the negative converter with current inversion (INIC), the capacitor and wires. $Y_{AA}$, $Y_{BB}$ and $Y_{CC}$ are the capacitance of the capacitor. $Y_{AA}^{'}$ and $Y_{CC}^{'}$ are the capacitance of the INIC for a capacitor. $A$ sub-chain, $B$ sub-chain and $C$ sub-chain are respectively connect by those electric devices. $Y_{AB}$ and $Y_{AB}^{'}$ being respectively the capacitance of the capacitor and INIC for a capacitor connect the $A$-subchain and $B$-subchain, while $Y_{BC}$ and $Y_{BC}^{'}$ connect the $B$-subchain and $C$-subchain. The on-site potential is obtained by grounding each node with suitable circuit devices.}
\label{fig4}
\end{figure*}

The singular energy spectrum can also be examined from a more general perspective. In ref. \cite{Yang21}, authors proposed a new concept of the auxiliary generalized Brillouin zone, which is obtained by solving the characteristic equation $f(\beta, E)=0$ when $E$ takes every value of the complex number field, and selecting $|\beta_{m}|=|\beta_{n}|$ ($\{m, n\}=\{1, 2, 3, 4\}$). However, as shown in Fig. \ref{fig3}(c), even though the auxiliary generalized Brillouin zone possesses all equality information of $f(\beta, E)=0$, it still can not recover the generalized Brillouin zone obtained from the open boundary eigenvalues. This further confirms the emergence of the singular energy spectrum.

Meanwhile, we analytically exhibit that the determinant of the open boundary Hamiltonian has the form $det\big[H_{OBC}-E_{OBC}; N\times N]=2^{-N}(V_{a}-E_{OBC})^{N}(V_{c}-E_{OBC})^{N}y(V_{b}, t_{b}, E_{OBC})$ with the unidirectional hopping being satisfied \cite{ennd32}, with N being the total number of the system size. Hence, $E_{OBC}=V_{a}$ and $E_{OBC}=V_{c}$ must be the solutions of $det\big[H_{OBC}-E_{OBC}; N\times N]=0$, and be the multifold exceptional points \cite{Delp36}$^{,}$\cite{Berg37}. In Fig. \ref{fig4}d, we exhibit the number of the open boundary eigenvalues of $E_{OBC}=V_{a}$ and $E_{OBC}=V_{c}$ with different system size. Explicitly, the number of degeneracy points is proportional to the system size.

Additionally, we know that the sensors have already penetrated every field of our life, in which one of the core parts is the transformation circuit, which enlarges the weak signals \cite{Dege36, Andr37, Wang38}. Interestingly, it has been displayed that the minute perturbations of the parameters will induce the mutations of the eigenvalues in our system. Moreover, Refs. \cite{Yin39, Ales40, Qin41, Koch42, Nikza43, Zhan44} also show that the higher-order exceptional points have advantages in sensors, and it is significant to seek higher-order exceptional points in various systems. Coincidentally, we also have shown that the emergence of the singular phenomenon is accompanied by the multifold exceptional point. Hence, our non-Hermitian systems may have huge applications in the sensors field.

\noindent $\textbf{Experimental implementation}$

\noindent There exist several physical incarnations that can be used to explore the non-Hermitian systems \cite{GW45, XY46, Hao47, Zhou48, Zou49, YT50, Lee51, Rus52, Flec53, Wu54, CA55, CW56, Ozaw57, Liu58, Zhen59}, including the cold atoms \cite{GW45, XY46, Hao47, Zhou48}, electrical circuits \cite{Zou49, YT50, Lee51, Rus52, Flec53, Wu54}, photonic and acoustic systems \cite{CA55, CW56, Ozaw57, Liu58, Zhen59}. Among them, electric circuits have been widely used because the circuit structure can be flexibly designed to facilitate integration and mass production. Here, we propose an experimental scheme to realize our system through electric circuits. The essential part of the electric system is the negative converter with current inversion (INIC), the impedance of which is changed from negative to positive with the orientation of the current being reversed, or vice versa, as shown in Fig. \ref{fig4}. We can achieve the system we considered by choosing appropriate impedances for these electric devices. Explicitly, the parameters of the devices are $Y_{A1}=V_{a}$, $Y_{Aj}=V_{a}+t_{a}$ $(1<j\leq N)$, $Y_{Aj}^{'}=-\gamma_{1}$, $Y_{AB}=-\frac{\gamma_{1}+\gamma_{2}}{2}$, $Y_{AB}^{'}=\frac{\gamma_{1}-\gamma_{2}}{2}$, $Y_{AA}=Y_{AA}^{'}=-\frac{t_{a}}{2}$, $Y_{B1}=Y_{BN}=V_{b}$, $Y_{Bj}=V_{b}+t_{b}$ $(1<j<N)$, $Y_{BC}=-\frac{\gamma_{3}+{\gamma_{4}}}{2}$, $Y_{BC}^{'}=\frac{\gamma_{3}-{\gamma_{4}}}{2}$, $Y_{BB}=-t_{b}$, $Y_{Cj}=V_{c}$ $(1\leq j<N)$, $Y_{CN}=V_{c}-t_{c}$, $Y_{Cj}^{'}=-(\gamma_{4}+t_{c})$, $Y_{CC}=-\frac{t_{c}}{2}$ and $Y_{CC}^{'}=\frac{t_{c}}{2}$.

\noindent $\textbf{Discussion}$

\noindent We have unraveled that the open boundary eigenstates can exhibit anomalous non-Hermitian skin effect: localized for topologically trivial point gap while extended for the nontrivial point gap, provided that the one-directional coupling amplitudes among the sub-chains are exhibited. Additionally, with the presence of the anomalous non-Hermitian skin effect, the open boundary eigenvalues will have wide distinction compared with the continuum bands. Notably, there exist multifold exceptional points in an open chain, and the degree of degeneracy will increase as long as the system size increases. The physical properties of the eigenvalues may have applications in the sensors field. Our results presented here exhibit the unique irrelevance of the non-Hermitian skin effect and the point gap, which can promote the development of the non-Bloch theory and the critical phenomenon in the non-Hermitian field. Meanwhile, the conclusions can be generalized to various non-Hermitian systems [such as four-band system\cite{ennd32}].

\noindent $\textbf{Methods}$

\begin{figure}[!htbp]
\includegraphics[width=4.27cm,height=3.4cm]{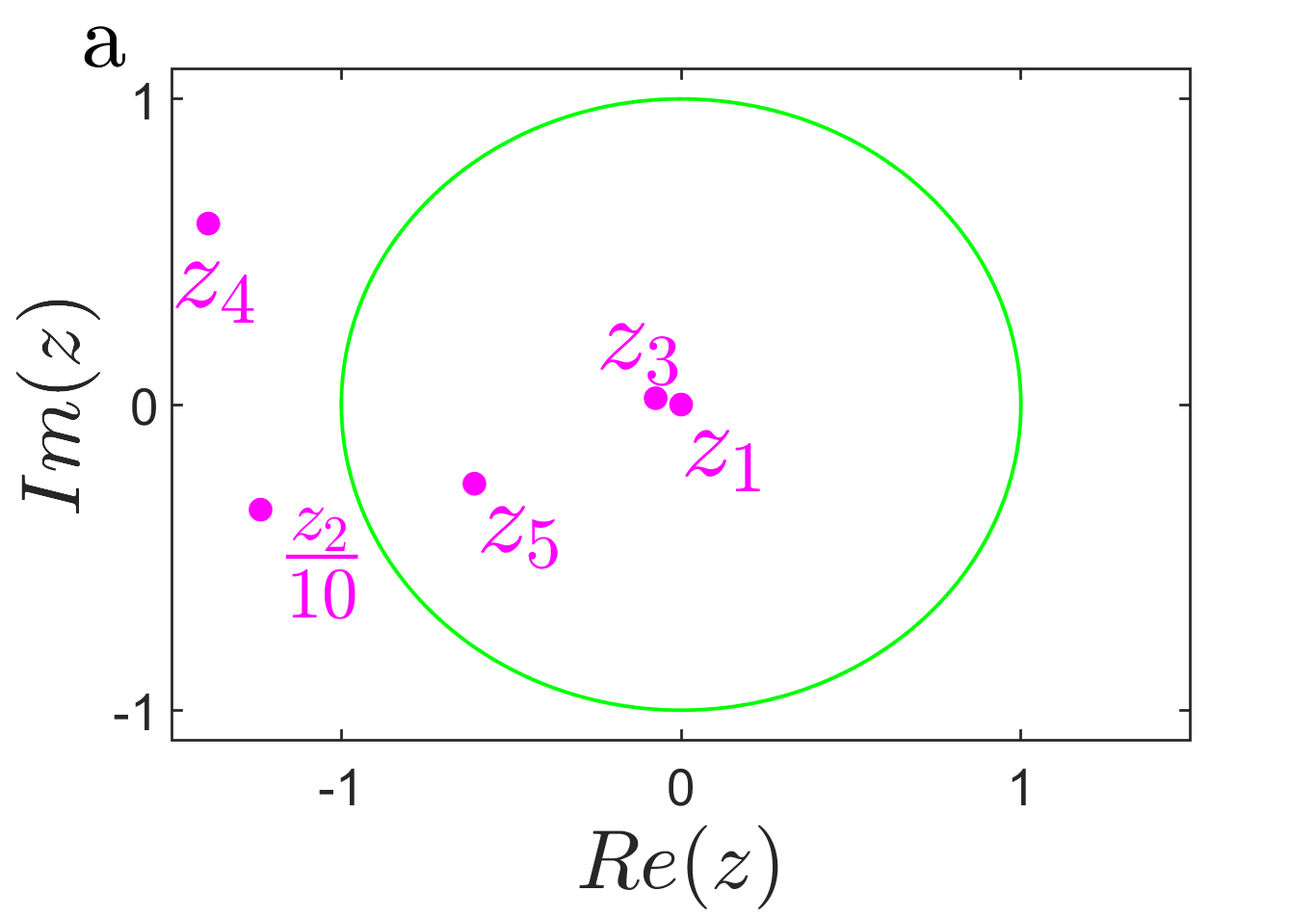}
\includegraphics[width=4.27cm,height=3.4cm]{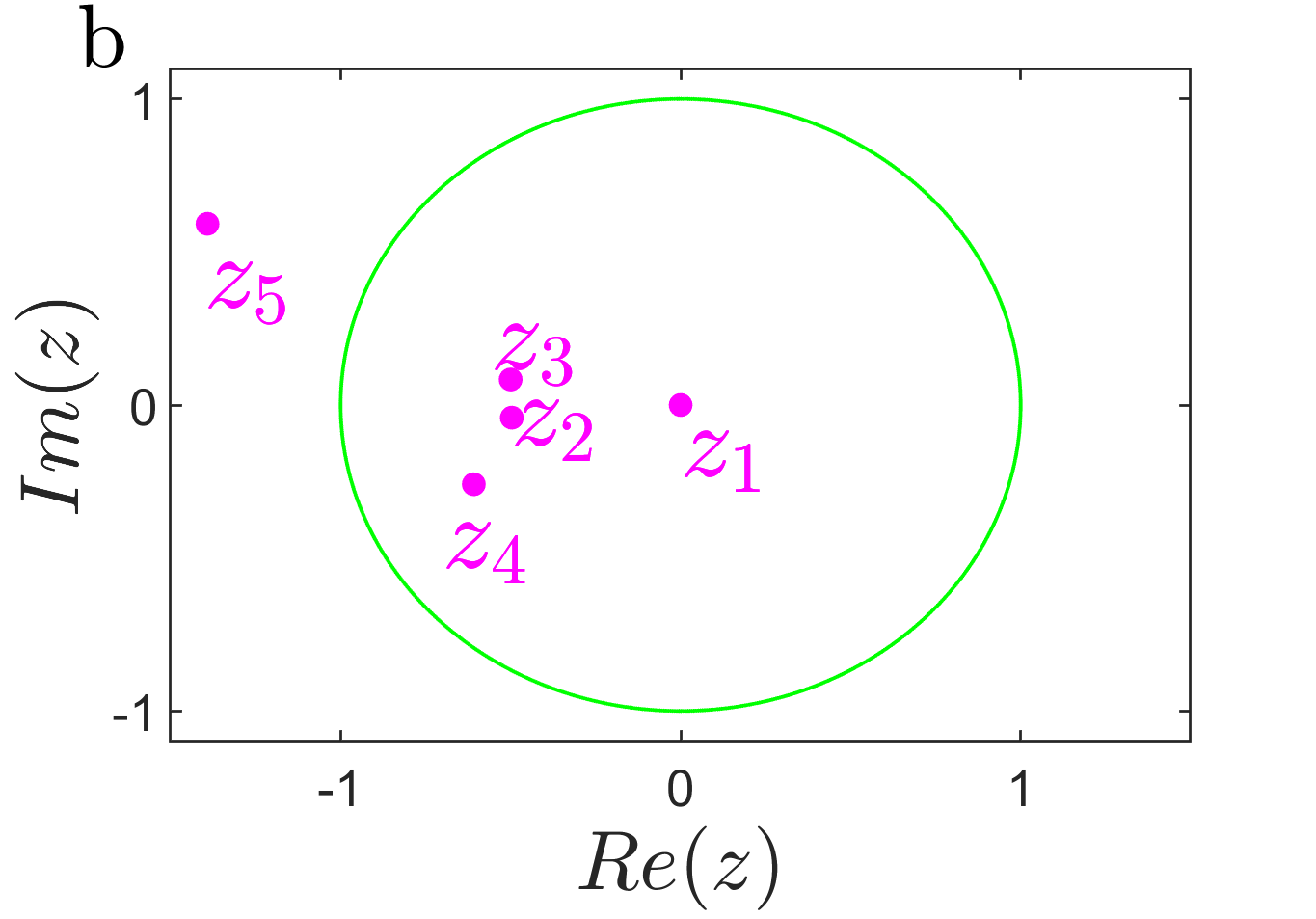}
\caption{\textbf{The distribution of the poles under the anomalous non-Hermitian skin effect situation}. \textbf{a} $z_{1}$, $z_{3}$, $z_{5}$ being encircled by the unit circle, while $z_{2}$ $z_{4}$ are outside the unit circle for a reference point $E_{ref}=\frac{1}{6}i$. Thus, the system corresponds to the trivial point gap. The common parameters are $\gamma_{1}=\gamma_{3}=0$, $t_{a}=t_{c}=8$, $t_{b}=\frac{1}{2}$, $\gamma_{2}=2$, $\gamma_{4}=1$, $V_{a}=V_{c}=\frac{3}{5}$ and $V_{b}=1$. \textbf{b} $z_{1}$, $z_{2}$, $z_{3}$ and $z_{4}$ being encircled by the unit circle, which induces the nonzero spectral winding number, or the nontrivial point gap, for a reference point $E_{ref}=\frac{1}{6}i$. The common parameters are $\gamma_{1}=\gamma_{4}=0$, $t_{a}=1$, $t_{b}=\frac{1}{2}$, $t_{c}=2$, $\gamma_{2}=2$, $\gamma_{3}=1$, $V_{a}=2$ and $V_{b}=V_{c}=1$.}
\label{fig55}
\end{figure}

\noindent $\textbf{The analytical expression of the point gap --- the}$
$\textbf{spectral winding number}$.  We discuss detailedly the spectral winding number based on the definition $W=\frac{1}{2\pi i} \int_{0}^{2\pi}dk\partial_{k}\ln\det[H(k)-E_{ref}]$ \cite{Zhang16}$^{,}$ \cite{Okum28, Gon29, Denn30, Yi31}, in which $E_{ref}$ is an arbitrary reference point. For our three-band system, the spectral winding number can be obtained quantitatively from the residue theorem via $e^{ik}\rightarrow z$ and $dk\rightarrow\frac{1}{iz}dz$.

Such as, one can see that five poles are spread on the complex plane for $E_{ref}=\frac{1}{6}i$ based on the definition, as shown in Fig. \ref{fig55}a. Further, three of them are surrounded by the unit circle, which induces $W=0$. Namely, the point gap is topologically trivial. [Precisely, the spectral winding $W\equiv0$ for any reference point $E_{ref}$ as long as $V_{a}=V_{c}$, $t_{a}=t_{c}$ and $\gamma_{1}\gamma_{2}=\gamma_{3}\gamma_{4}$ \cite{ennd32}]. In addition, fives poles of order one are distributed on the complex plane for a given $E_{ref1}=\frac{1}{6}i$ in Fig. \ref{fig55}b, where four of them are contributed to the spectral windings \cite{ennd32} and $W=1$ can be obtained, i.e., the point gap is topologically nontrivial.

\bibliography{refpar}

\noindent $\textbf{Acknowledgements}$

\noindent This work was supported by NSFC under grants No.11874190. NSFC No. 12047501. National Key R$\&$D Program of China under grants No. 2021YFA1400900, 2021YFA0718300, 2021YFA1402100. Fundamental Research Funds for the Central Universities, China, No. FRF-TP-22-098A1

\end{document}